\documentclass[review,number,sort&compress]{elsarticle}

\usepackage[T1]{fontenc} 
\usepackage[utf8]{inputenc}

\usepackage{graphicx,epstopdf}
\usepackage{caption}
\usepackage{subcaption}
\usepackage{amssymb}
\usepackage{amsmath}
\usepackage[section]{placeins}
\usepackage{siunitx}
\usepackage{color}

\journal{Nuclear Instruments and Methods in Physics Research Section A}

\begin{document}

\begin{frontmatter}

\title{Envelope Dynamics and Stability with non-linear Space-Charge Forces}

\author{Michael Holz\corref{cor1}}
\ead{michael.holz@physics.uu.se}
\author{Volker Ziemann}

\address{Uppsala University, Lägerhyddsvägen 1, Box 516, Uppsala, Sweden}

%
\cortext[cor1]{Corresponding author}

%

\begin{abstract}
We developed a model to calculate the stability of Gaussian beam distributions with non-linear space-charge forces in the presence of random and skew-quadrupole errors. The effect of the space-charge force on the beam matrix is calculated analytically including full cross-plane coupling in 4D phase space, which allows us to perform fast parameter studies. For stability analysis, we find the fixed points of the beam including the space-charge forces and construct a Jacobi-matrix by slightly perturbing the periodic solution. The stability of envelope oscillations is inferred by eigenvalue analysis. Furthermore, we employ envelope tracking as a complementary method and compare the results of the eigenvalue analysis with FFT data from the tracked envelope. The non-linearity of the space-charge force in combination with lattice errors and beam coupling opens up for envelope-lattice resonances and envelope coupling resonances. Hitting these resonances leads to envelope blow-up, causing an effective beam mismatch. Therefore, we finally examine the effect of beam mismatch on the envelope tune-shift and its stability.
\end{abstract}

\begin{keyword}
space-charge, beam stability, quadrupole error, skew quadrupole, stop-band
\end{keyword}

\end{frontmatter}


\section{Introduction}

With the development of high-intensity linacs and accumulator rings, the space-charge force and its detrimental effects on beam dynamics became one of the main study subjects. In terms of beam dynamics, it causes tune-shifts which have the potential to drive the beam into resonances. Furthermore, it contributes to beam halo formation and emittance blow-up, which cause beam losses and a subsequent generation of radiation and heat. A thorough understanding of the underlying dynamical mechanisms is therefore essential.

In order to describe collective space-charge effects, it is common to look at the evolution of the beam envelope. Established techniques are based on envelope equations or numerical methods like multi-particle tracking from which rms values are calculated. Envelope equations are a computationally fast approach to beam envelope dynamics with space-charge and were developed by Sacherer \cite{Sacherer:322516}. They use linear approximations for the space-charge force and can be represented by beam matrices, containing the second moments of a beam distribution. They may also include linear coupling terms in two transverse dimensions. In \cite{Gluckstern:1998zz}, these envelope equations were used to create stationary, field-generating beam distributions, which are combined with tracking of spectator particles and particle-in-cell (PIC) codes to study the development of beam halo, in particular, parametric resonances. PIC codes are usually computationally expensive, depending on grid sizes and the amount of interactions taken into account. In \cite{PhysRevSTAB.18.024202}, the authors use tracking of single particles with space-charge kicks derived from the potential of an upright Gaussian distribution, which itself is derived from the rms quantities of the single particles. Through this calculation sequence, the number of applied space-charge kicks is limited if simulations are to be completed in a limited time-frame. In \cite{Struckmeier:144594,Li:2015PoP}, envelope stability was examined by means of eigenvalue analysis and based on envelope equations with linear forces and complemented with particle tracking.

Our model is based on constructing a map for the transverse coupled beam matrix that includes the space-charge forces, as described by the Bassetti-Erskine formula \cite{Bassetti:122227,Ziemann:1991sb}. We then proceed to determine the fixed points of the map and perform a linear stability analysis around them. From the eigenvalues of the linearized map, we infer the stability of configurations. Finding the equilibrium beam non-perturbatively first and then perturbing it to identify the coherent eigenfrequencies resembles the analysis in \cite{doi:10.1142/5761}.

The model we developed is presented Section \ref{sec:model} and complements the established techniques. It tracks the second moments of Gaussian distributions including non-linear forces without the need of intermediate rms value calculation, effectively combining positive aspects from both analytical and numerical approaches. In addition, the model's ability to treat cross-plane coupling enables us to simulate cases that include coupled beams and lattices. 
The section also includes a description of the ring lattice and of the method we use for envelope stability analysis. In section \ref{sec:enstabdyn}, we present and discuss our findings of envelope dynamics and stability. Finally, we examine the effect of beam mismatch on the coherent envelope tune-shift in section \ref{subsec:mismatch}.

\section{Space Charge Modeling and Methods}
\label{sec:model}


In \cite{Bassetti:122227}, Bassetti and Erskine derived a closed, analytic expression for the electric field of upright Gaussian beams.
Their formula was generalized in \cite{Ziemann:1991sb} to include all correlations within the 4D transverse phase space. The formula then reads

\begin{equation}
\begin{split}
f_{3}+if_{1}=F_{0}\left(x_{1},x_{3},\sigma\right) = \frac{\sqrt{\pi}}{\sqrt{2(\sigma_{11}-\sigma_{33}+2i\sigma_{13})}} \left[ w\left( z_{1}\right) - \exp\left[-\frac{1}{2} \sum\limits_{l,m=1,3}\sigma^{-1}_{lm}x_{l}x_{m}\right] w\left(z_{2}\right)\right]
\end{split}
\label{eq:F0}
\end{equation}
where $\vec{x}=\: \left(x_{1},x_{2},x_{3},x_{4}\right)$ contains the positions and angles of particles and $w(z)$ is the complex error function \cite{Abramowitz}. The sum in the exponent in the above equation contains only the indices \num{1} and \num{3} of the position coordinates, since the self-force does not depend on the angles. The arguments of the complex error functions are given by

\begin{equation}
	\begin{aligned}
		z_{1} &= \frac{x_{1}+ix_{3}}{\sqrt{2(\sigma_{11}-\sigma_{33}+2i\sigma_{13})}}&=\sum\limits_{k} a_{k}x_{k} = a_{1}x_{1}+a_{3}x_{3}\\
		z_{2} &= \frac{(\sigma_{33}-i\sigma_{13})x_{1}+i(\sigma_{11}+i\sigma_{13})x_{3}}{\sqrt{\sigma_{11}\sigma_{33}-\sigma^{2}_{13}}\sqrt{2(\sigma_{11}-\sigma_{33}+2i\sigma_{13})}} &=\sum\limits_{k} b_{k}x_{k} = b_{1}x_{1}+b_{3}x_{3}.
		\end{aligned}
\label{eq:z}
\end{equation}
We interpret the kick to the beam envelope represented by the second moments of a Gaussian distribution by its force as an average kick received by all of its constituent particles. Through this averaging, we construct a mapping for the space charge force for the ten beam matrix elements

\begin{equation}
	\begin{array}{l l l l l l}
		\bar{\sigma}_{11} &=& \langle\bar{x}_{1}\bar{x}_{1}\rangle= \langle x_{1}x_{1}\rangle=\sigma_{11}\\
		\bar{\sigma}_{12}	&=&	\langle\bar{x}_{1}\bar{x}_{2}\rangle=	\langle x_{1}(x_{2}+\tilde{K}f_{1})\rangle=\sigma_{12}+ \tilde{K} \langle x_{1}f_{1}\rangle\\
		\bar{\sigma}_{13}	&=&	\langle\bar{x}_{1}\bar{x}_{3}\rangle=	\langle x_{1}x_{3}\rangle=\sigma_{13}\\
		\bar{\sigma}_{14}	&=& \langle\bar{x}_{1}\bar{x}_{4}\rangle=	\langle x_{1}(x_{4}+\tilde{K}f_{3})\rangle=\sigma_{14}+ \tilde{K} \langle x_{1}f_{3}\rangle\\
		\bar{\sigma}_{22}	&=&	\langle\bar{x}_{2}\bar{x}_{2}\rangle=	\langle (x_{2}+\tilde{K}f_{1})(x_{2}+\tilde{K}f_{1})\rangle=\sigma_{22}+ \tilde{K} \langle 2x_{2}f_{1}\rangle+ \tilde{K}^{2} \langle f^{2}_{1}\rangle\\
		\bar{\sigma}_{23}	&=&	\langle\bar{x}_{2}\bar{x}_{3}\rangle=	\langle x_{3}(x_{2}+\tilde{K}f_{1}\rangle=\sigma_{23}+ \tilde{K} \langle x_{3}f_{1}\rangle\\
		\bar{\sigma}_{24}	&=&	\langle\bar{x}_{2}\bar{x}_{4}\rangle=	\langle (x_{2}+\tilde{K}f_{1})(x_{4}+\tilde{K}f_{3})\rangle=\sigma_{24}+ \tilde{K} \langle x_{2}f_{3}\rangle+ \tilde{K} \langle x_{4}f_{1}\rangle+ \tilde{K}^{2} \langle f_{1}f_{3}\rangle\\
		\bar{\sigma}_{33}	&=&	\langle\bar{x}_{3}\bar{x}_{3}\rangle=	\langle x_{3}x_{3}\rangle=\sigma_{33}\\
		\bar{\sigma}_{34}	&=&	\langle\bar{x}_{3}\bar{x}_{4}\rangle=	\langle x_{3}(x_{4}+\tilde{K}f_{3})\rangle=\sigma_{34}+ \tilde{K} \langle x_{3}f_{3}\rangle\\
		\bar{\sigma}_{44}	&=&	\langle\bar{x}_{4}\bar{x}_{4}\rangle=	\langle (x_{4}+\tilde{K}f_{3})(x_{4}+\tilde{K}f_{3})\rangle=\sigma_{44}+ \tilde{K} \langle 2x_{4}f_{3}\rangle+ \tilde{K}^{2} \langle f^{2}_{3}\rangle,
	\end{array}
	\label{eq:newEl}
\end{equation}
where $\tilde{K}=\: K\Delta\ell$ and the beam perveance is given by

\begin{equation}
K = \frac{2Nr_{0}}{\sqrt{2\pi}\sigma_{z}\beta^{2} \gamma^{3}}.
\end{equation}
$N$ is the number of protons, $r_{0}$ their classical particle radius, and $\sigma_{z}$ the longitudinal beam size. The space-charge force is calculated in each longitudinal lattice slice of length $\Delta\ell$. The angle brackets in eq. \ref{eq:newEl} denote the averaging over the particle distribution given by

\begin{equation}
\Psi(\vec{x}) = \frac{1}{\left(2\pi\right)^{2}\sqrt{\det\sigma}} e^{-\frac{1}{2}\sum\limits^{4}_{l,m=1}\sigma^{-1}_{lm}x_{l}x_{m}},
\end{equation}
where $\sigma$ is the full 4D beam matrix, which is identical to the beam matrix used in calculating the electric field. The averages appearing in Eq. \ref{eq:newEl} can therefore be written as

\begin{equation}
\begin{aligned}
\langle x_{k}F_{0}\rangle &= \langle x_{k}(f_{3}+if_{1})\rangle =\int\limits_{-\infty}^{\infty} d^{4}x \ \Psi(\vec{x})\:  x_{k} F_{0}(x_{1},x_{3},\sigma) \\
\langle F_{0}\bar{F_{0}}\rangle &=\langle f_{3}^{2}\rangle = \langle f_{1}^{2}\rangle =  \int\limits_{-\infty}^{\infty} d^{4}x\  \Psi(\vec{x})F_{0}\bar{F_{0}}\\
\langle f_{1}f_{3}\rangle&=\:0 .
\end{aligned}
\label{eq: IntSol}
\end{equation}
For these integrals, we derived fully analytic solutions in \cite{Holz1160961}. We are thus able to perform envelope tracking  to apply non-linear space-charge kicks in a beam matrix formalism. The inhomogeneous terms in Eq. \ref{eq:newEl} are collected in a matrix $\left(\sigma_{sc}\right)_{ij}= \bar{\sigma}_{ij}-\sigma_{ij}$ and added to beam matrix of the preceding step. We point out that Gaussians are not self-consistent distributions under the influence of non-linear space-charge forces. The introduced discrepancies are, however, small in the considered regime of operation with a maximum incoherent tune-shift on the order of \num{0.5}.

\begin{figure}[hbt]
\centering
\includegraphics[width=0.6\textwidth]{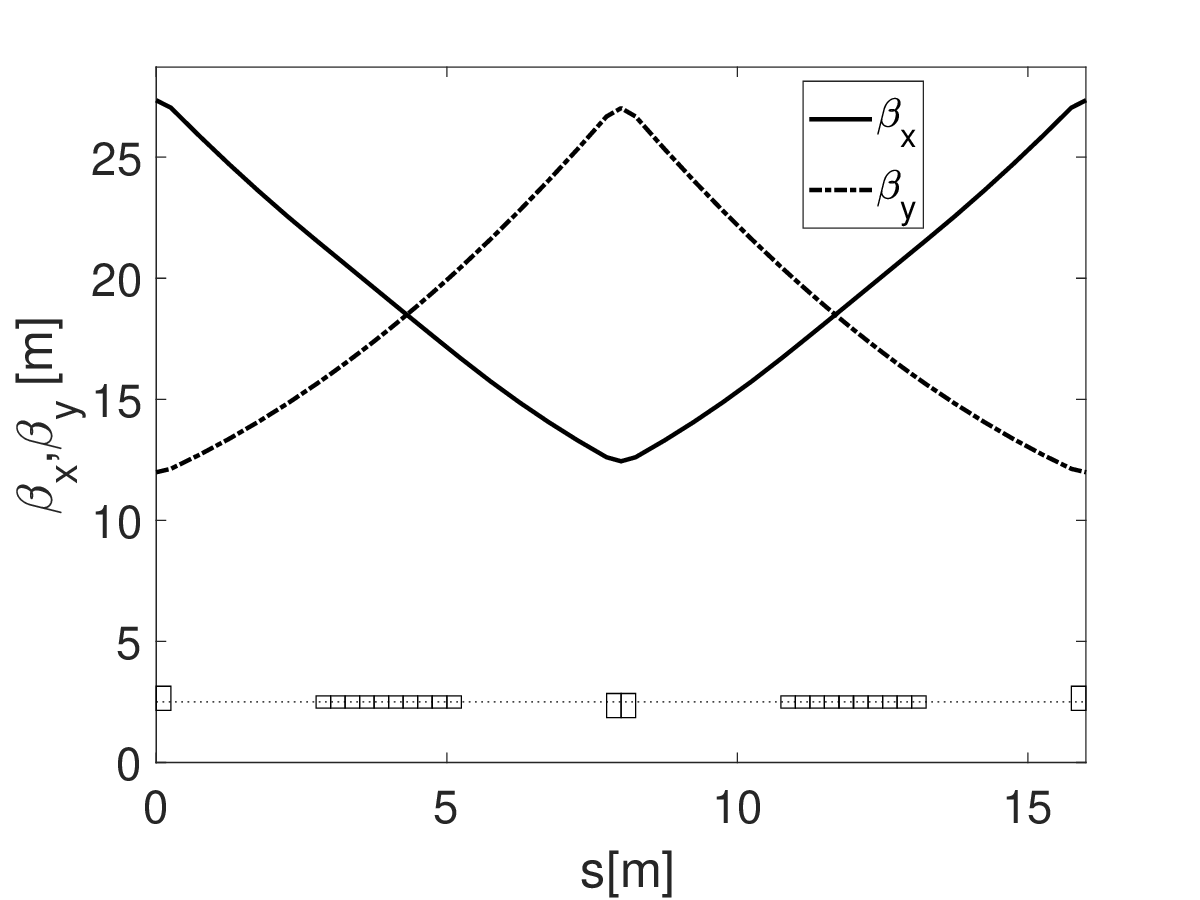}
  \caption{$\beta$-functions in one cell with length \SI{18}{\metre}. The whole ring consists of \num{18} cells.}
  \label{fig:lattice}
\end{figure}

In order to apply the space-charge kick in a symmetric way, we split the slices with transfer matrix $R$ into halves $\hat{R}$ such that $R=\:\hat{R}^{2}$ and propagate the beam via

\begin{equation}
\bar{\sigma}= \hat{R}\left( \hat{R}\sigma \hat{R}^{T} + \sigma_{sc} \right)\hat{R}^{T}
\end{equation}
through each slice. The lattice we use throughout this work consists of a ring with \SI{324}{\metre} circumference, comprised of \num{18} identical FODO-cells. We have  reconstructed this lattice from information in \cite{PhysRevSTAB.18.024202}. The qualitative configuration of one cell is depicted in Fig. \ref{fig:lattice}. The quadrupoles have a length of \SI{50}{\centi\metre} with k-values of \SI{0.1795}{\per\metre\squared} and \SI{-0.2071}{\per\metre\squared} for the focusing and defocusing quadrupoles, respectively. The bending magnets have a length of \SI{3.5}{\metre} and a \SI{10}{\degree} bending radius. The tunes of the lattice, purely defined by the machine elements, are $Q_{x}=\:\num{2.6}$ horizontally and $Q_{y}=\:\num{2.95}$ vertically. We set the beam energy to \SI{7}{\mega\electronvolt}. 

We divide the ring in \num{6480} equidistant slices with length $\Delta \ell=\:$ \SI{5}{\centi\metre}. For \num{1024} turns in the ring, that amounts to more than \num{6.5} million slices. In each slice, the space-charge kick is applied to the beam. The analytic solutions of the non-linear space-charge kick enables us to complete such a simulation for one value of the particle density $\frac{dN}{dz}$ within a few minutes on a desktop computer.

We find the zero-current periodic solution via the parametrization shown in the appendix of \cite{PhysRevSTAB.18.072801}. If the beam enters the lattice with its zero-current periodic solution, it experiences an injection mismatch due to the sudden presence of space-charge forces. In order to exclude the injection mismatch from our simulation studies, we find the periodic solution for the beam matrix including the non-linear space-charge force. For that, we propagate the beam one turn through the lattice and search the minimum of the cost function

\begin{equation}
\chi^{2}\left(\hat{\sigma}^{in}\right)= \sum_{i}\left(\hat{\sigma}^{out}_{i}-\hat{\sigma}^{in}_{i}\right)^{2},
\end{equation}
where $\hat{\sigma}$ is a column vector with a selection from the ten independent beam matrix elements.
In the reduced uncoupled case, our input parameters are $\hat{\sigma}=\: \left[\sigma_{11},\sigma_{12},\sigma_{22},\sigma_{33},\sigma_{34},\sigma_{44}\right]^{T}$.

The stability of the beam matrix and its phase-advance are determined by systematically perturbing the periodic solutions about a small value $\pm\frac{\Delta_{i}}{2}$ and observe the average change $\langle\epsilon_{i,N}\rangle$ in all other parameters after one turn trough the lattice. Each parameter change yields a column vector with the change ratio $\frac{\epsilon}{\Delta}$ from which we construct a Jacobi-matrix $J$. The principle of this construction is outlined in Eq. \ref{eq:jacobi} for all ten independent beam parameters.

\begin{equation}
\begin{pmatrix}\hat{\sigma}_{11}\pm\frac{\Delta_{1}}{2}\\\hat{\sigma}_{12}\\ \vdots \\ \hat{\sigma}_{44} \end{pmatrix} \stackrel{one\: turn}{\rightarrow} \begin{pmatrix} \hat{\sigma}_{11}+\langle\epsilon_{1}\rangle \\ \hat{\sigma}_{12}+\langle\epsilon_{2}\rangle \\ \vdots\\ \hat{\sigma}_{44}+\langle\epsilon_{10}\rangle  \end{pmatrix}; \quad J= \begin{pmatrix} \frac{\langle\epsilon_{1}\rangle}{\Delta_{1}} & \frac{\langle\epsilon_{1}\rangle}{\Delta_{2}} & \dots & \frac{\langle\epsilon_{1}\rangle}{\Delta_{10}} \\  \frac{\langle\epsilon_{2}\rangle}{\Delta_{1}} & \frac{\langle\epsilon_{2}\rangle}{\Delta_{2}} & \dots & \frac{\langle\epsilon_{2}\rangle}{\Delta_{10}} \\ \vdots & \vdots & \ddots& \vdots \\ \frac{\langle\epsilon_{10}\rangle}{\Delta_{1}} & \frac{\langle\epsilon_{10}\rangle}{\Delta_{2}} & \dots & \frac{\langle\epsilon_{10}\rangle}{\Delta_{10}} \end{pmatrix}
\label{eq:jacobi}
\end{equation}
 Its eigenvalues contain one complex-conjugate pair for each oscillating mode. Additionally, two of the eigenvalues are real and unity. They are the invariants of the system, which accounts for the beam emittance.  The phase advances are extracted from angle of the complex eigenvalue. 

Since the system's dynamic variable is the beam size, the angles of the eigenvalues contain twice the phase-advance of the lattice in the zero-current limit.

In the following section, we present and discuss the results of the stability analysis for test cases for undisturbed lattices, quadrupole errors and skew quadrupoles.

\section{Envelope Stability and Dynamics}
\label{sec:enstabdyn}

In this section, we examine the stability of the coherent envelope modes by means of eigenvalue analysis under the presence of space-charge forces.
In our results, we show the fractional part of the eigentunes $Q$ as a function of charge density, and a corresponding stability plot with eigentunes and the absolute eigenvalues in order locate resonances and instabilities. We assume a transverse emittance of \SI{1}{\milli\metre \milli\radian} in both planes. For the largest tested charge density, the coherent envelope tune-shift is \num{-0.3} and the maximum incoherent tune-shift is \num{-0.52}.


We start the analysis with an undisturbed lattice. Then, we introduce single and later random quadrupole errors, where we expect the coherent envelope modes to react to the created stop-bands. The full cross-plane coupling enables us to examine simulation cases with single or randomly applied roll angles to quadrupoles. Additionally, the space-charge force in our model couples the coherent eigenmodes, which enables resonances between them \cite{Hofmann:2017isl}.


\subsection{Envelope Dynamics in the undisturbed Lattice}

\begin{figure}[hbt]
    \centering
    \begin{subfigure}{0.49\textwidth}
        \centering
        \includegraphics[width=\textwidth]{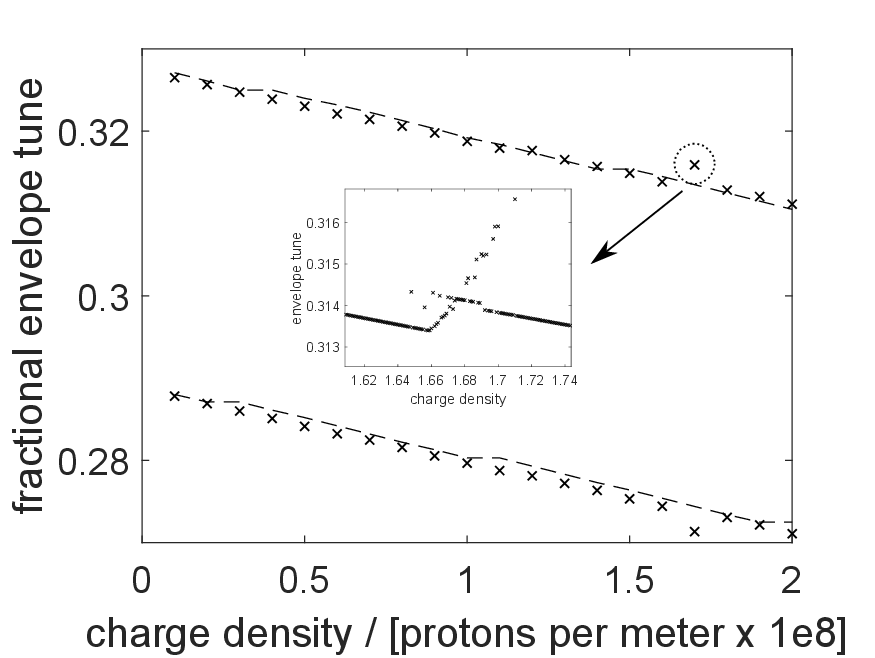}
        \caption{}
				\label{fig:undistOneCellFFT}
    \end{subfigure}%
     ~
    \begin{subfigure}{0.49\textwidth}
        \centering
        \includegraphics[width=\textwidth]{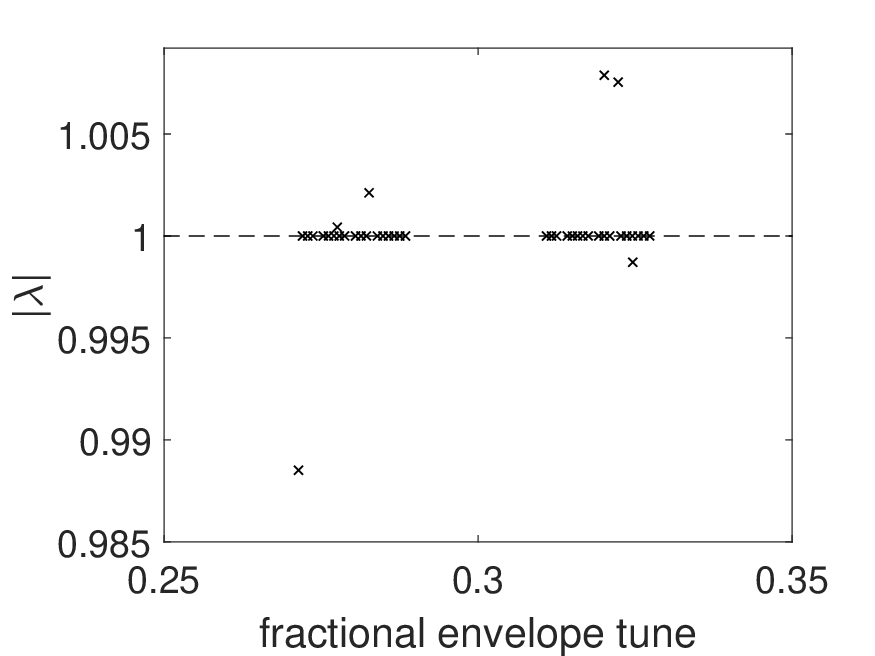}
        \caption{}
				\label{fig:undistOneCellEig}
    \end{subfigure}
    \caption{a) Comparison between coherent envelope tunes obtained by eigenvalue analysis (crosses) and Fourier analysis of tracked envelopes (dashed lines). b) Eigenvalue moduli versus fractional envelope tune of the two envelope modes for a single cell.}
		\label{fig:1cell}
\end{figure}

Before we examine the envelope stability in the test ring, we test the functionality of the eigenvalue analysis using one cell and show the results in Figure \ref{fig:1cell}. A stable envelope mode is characterized by complex pairs of conjugate eigenvalues $\lambda$ with modulus $\left\lvert\lambda\right\rvert=\:\num{1}$. The results with one cell deviate from unity by less than \num{5e-4} in both modes, which is due to the accumulation of numerical noise. We confirm the functionality of the eigenvalue analysis by comparing the tunes obtained from the eigenvalues and by Fourier analysis of the envelope, tracked over \num{1024} revolutions. The coherent envelope tunes follow the expected linear shift as the beam current increases and the tunes of both methods agree well. There are small deviations visible around the charge density \num{1.7e8} protons per meter, where the moduli increase. Repeating the simulation with more data points, as seen in the inset reveals, that the tunes fulfill  the $\num{5}Q_{1}+\num{2}Q_{2}\approx\:2$ resonance and that it is indeed a physical effect. Exchange between the transverse planes is intrinsically possible in our space-charge model through the quadratic terms of the map in Eq. \ref{eq:newEl} and occurs independent of lattice errors.

\begin{figure}[bt]
    \centering
    \begin{subfigure}{0.49\textwidth}
        \centering
\includegraphics[width=\textwidth]{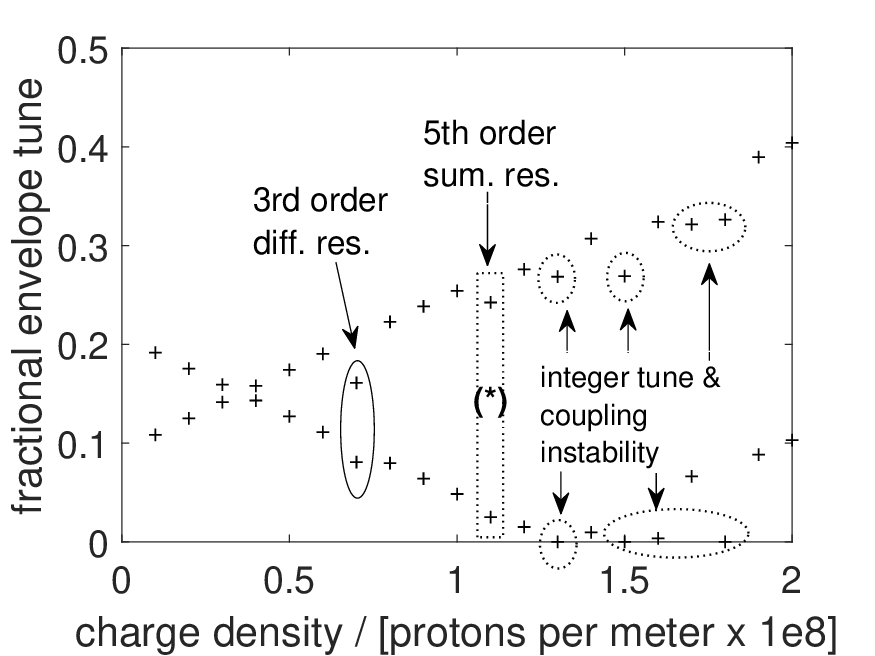}
        \caption{}
				\label{fig:EigvalsringQvsdNdz}
    \end{subfigure}%
     ~
    \begin{subfigure}{0.49\textwidth}
        \centering
        \includegraphics[width=\textwidth]{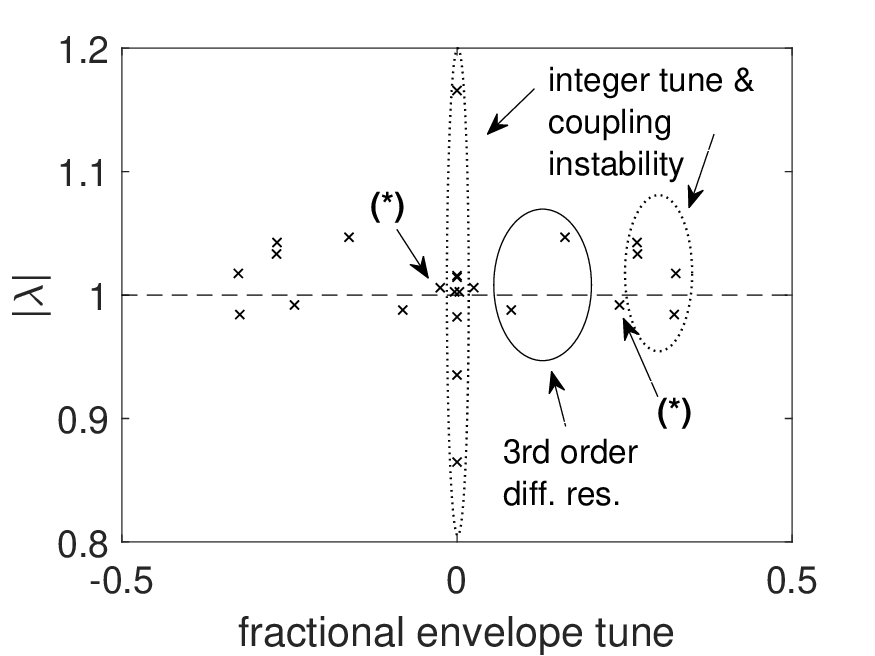}
        \caption{}
				\label{fig:EigvalsringEIG}
    \end{subfigure}
		    \caption{a) Fractional coherent envelope tune as function of charge density for the ring. b) Eigenvalue moduli of unstable envelopes. Instability due to integer tune-crossing and space-charge coupling (dotted ellipse), 3rd order difference resonance (solid ellipse), and 5th order sum resonance ((*)).}
		\label{fig:Eigvalsring}
\end{figure}

Repeating the simulations with the complete ring, we find the tunes of the ring, in combination with the tune-shift, causes one mode to be shifted towards an integer tune and finally, across it. Figure \ref{fig:Eigvalsring} shows the shifted tunes and the stability analysis for this case. The mode, which crosses the integer tune, exhibits instability with moduli of the eigenvalues up to \num{1.17}. This indicates that non-linear space-charge forces cause instability around the integer tune, even in the absence of lattice errors. In \cite{PhysRevSTAB.18.024202}, crossing the integer tune leads to an increase in tune spread, because the rms envelope is calculated from single particle coordinates. These particles can rearrange themselves in the case of light instabilities. That stands in contrast to our model, where instability directly leads to an increase of the envelope. Thus, the instability in the presented results is similar to quadrupole error-created stop-bands where the space-charge force is interpreted as adding a defocusing quadrupole in both planes in each slice of the lattice. Technically, it is half-integer stop-bands which are created. However, these also fall on integer tunes since they repeat in half-integer tune intervals. During integer tune crossing, the other transverse mode exhibits instability as well. We attribute this to the intrinsic space-charge coupling. This observation is consistent with \cite{PhysRevSTAB.18.024202}, where the tune spread also increases for that mode. Furthermore, we find third-order difference resonance and a fifth-order sum resonance, which are space-charge driven.

\FloatBarrier
\subsection{Envelope Dynamics with single and random Quadrupole Errors}

Now, we examine the envelope stability and tunes for single and random quadrupole errors. They break the symmetry of the lattice and create half-integer stop-bands. At the end of this section, we provide an analysis of how the stop-band contributes to emittance growth and thus, blowing up the envelope.

\begin{figure}[bt]
    \centering
    \begin{subfigure}{0.49\textwidth}
        \centering
\includegraphics[width=\textwidth]{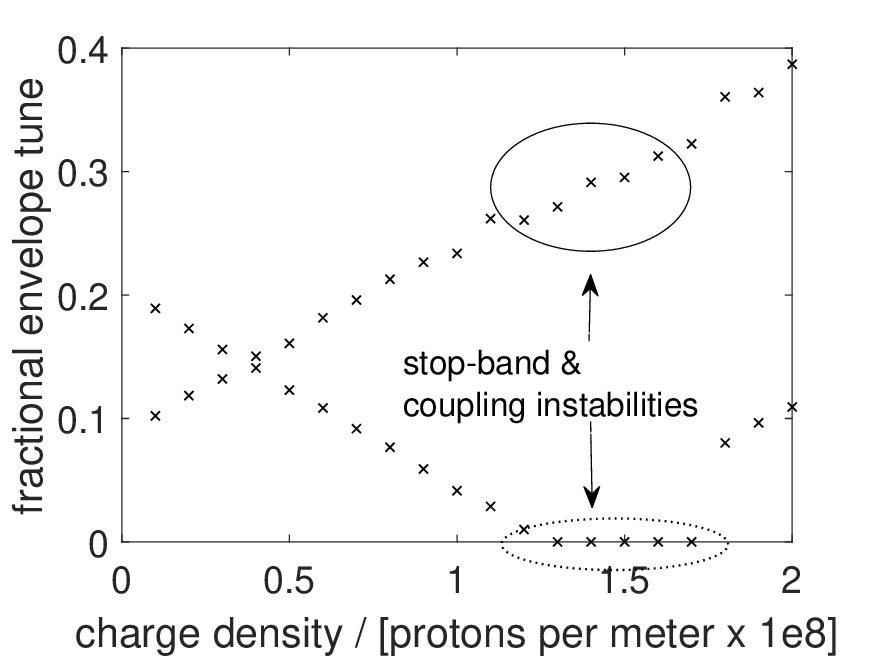}
        \caption{}
				\label{fig:oneErr1pcQvsdNdz}
    \end{subfigure}%
     ~
    \begin{subfigure}{0.49\textwidth}
        \centering
        \includegraphics[width=\textwidth]{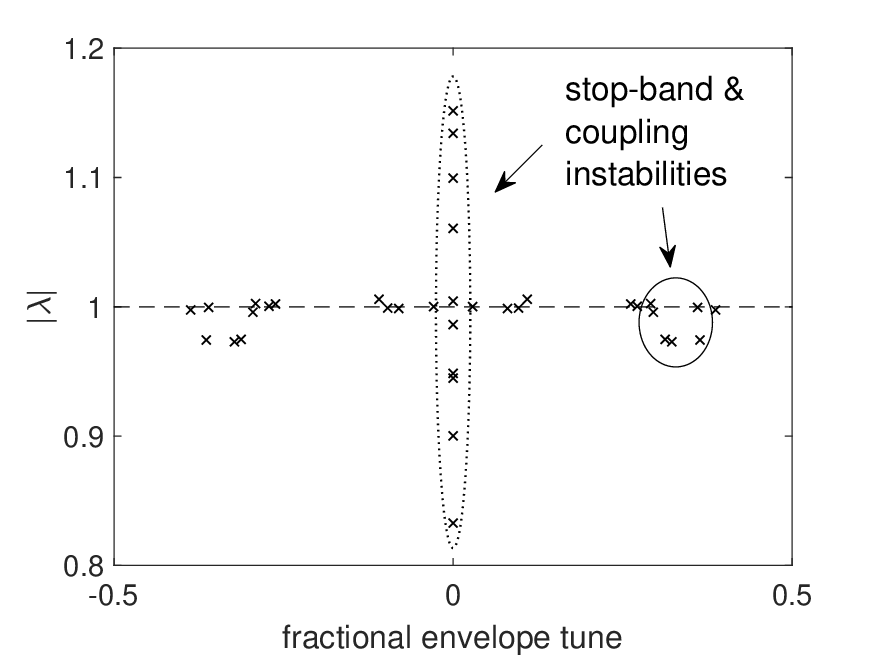}
        \caption{}
				\label{fig:oneErr1pcEig}
    \end{subfigure}
		    \caption{a) Coherent envelope tune-shift, where one mode crosses the integer tune at around a charge density of \num{1.3e8} protons per meter. The other transverse mode shows exhibits instability for the respective charge densities as well. b) Moduli of eigenvalues that show instability, with eigenvalues on the stop-band (dotted ellipse) and their corresponding coupling instabilities in the opposite plane (solid ellipse).}
		\label{fig:oneErr1pc}
\end{figure}

We start by slightly increasing the nominal strength of a single quadrupole by \SI{1}{\percent}. The shift of the lattice tune due to this error is $\Delta Q\approx \num{1e-3}$. Figure \ref{fig:oneErr1pc} shows the shifted envelope tunes and the corresponding moduli of unstable eigenvalues. In comparison to the undisturbed lattice, the eigenvalues of the integer-crossing mode are now locked consecutively on the integer tune (dotted ellipse). The stop-band has a certain width in tune, which means the beam envelope is driven into instability if its tune falls into the stop-band. While the stop-band in one plane is hit, we find clear coupling instabilities of the opposite plane (solid ellipse). Since the beam sizes of both planes influence each other through space-charge, a rapid growth and beta-beating in one plane induces instability in the other plane.

\begin{figure}[bt]
    \centering
    \begin{subfigure}{0.49\textwidth}
        \centering
\includegraphics[width=\textwidth]{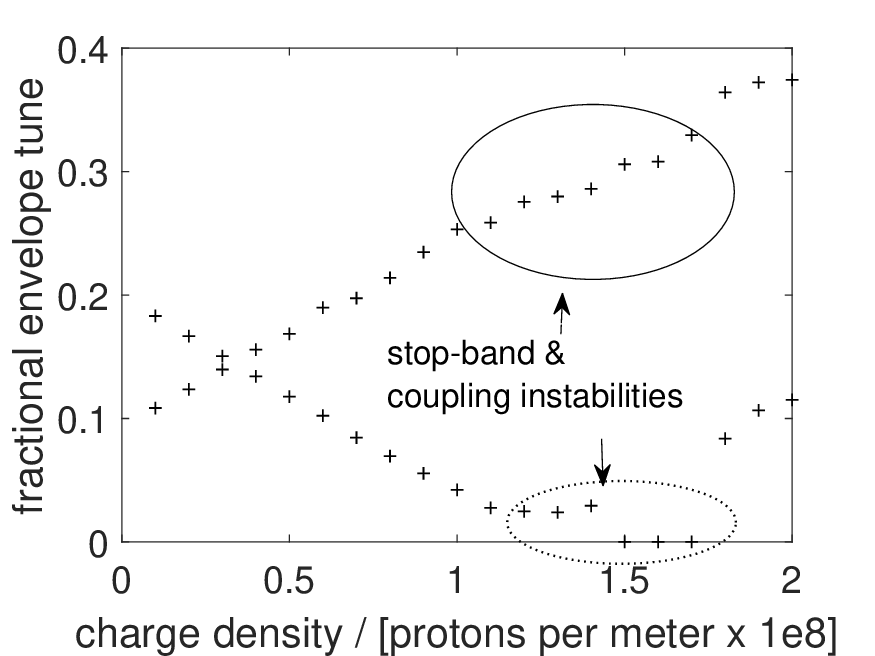}
        \caption{}
				\label{fig:randErr1pcdQvsdNdz}
    \end{subfigure}%
     ~
    \begin{subfigure}{0.49\textwidth}
        \centering
        \includegraphics[width=\textwidth]{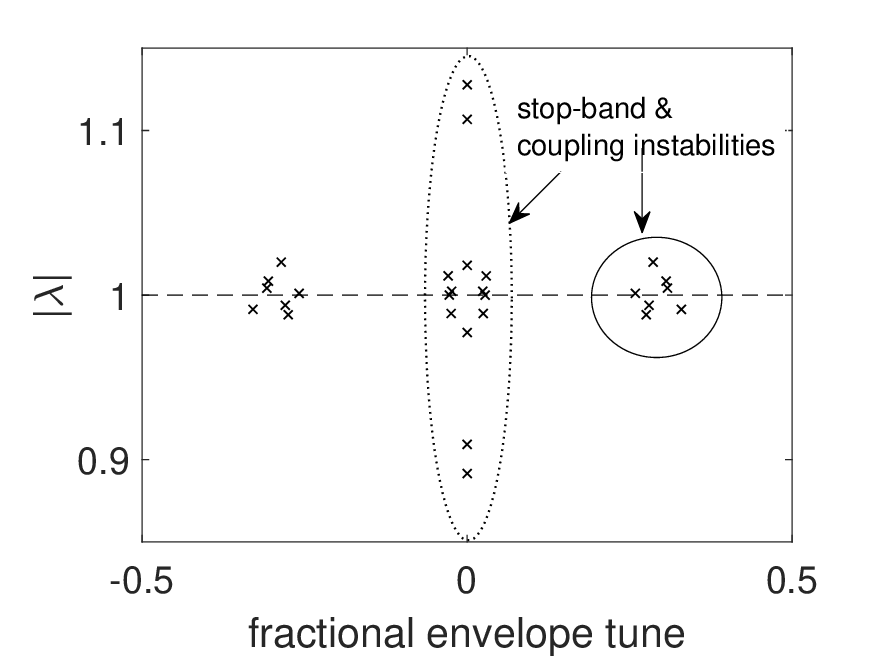}
        \caption{}
				\label{fig:randErr1pcdEig}
    \end{subfigure}
		    \caption{a) Tunes as a function of charge density for random quadrupole error variations of $\pm$\SI{1}{\percent}, applied to all quadrupoles. Instability due to one mode hitting the stop-band (dotted ellipse) leads to coupling instabilities in the other mode (solid ellipse). b) Moduli of the eigenvalues for the unstable modes.}
		\label{fig:randErrEigs}
\end{figure}

In the next simulation, we apply a \SI{1}{\percent} random variation to the strength of {\em all} quadrupoles and show the results in Fig. \ref{fig:randErrEigs}. The errors are sampled from a uniform distribution. The lattice tune has shifted with $\Delta Q= \num{5e-3}$, which is compatible with adding random perturbations in quadrature.

Regarding envelope stability, the number of unstable moduli in the vicinity of the integer tune is the same compared to the single \SI{1}{\percent} quadrupole error. The exact width and strength of the stop-band with random quadrupole errors strongly depends on the individual seed.

In Figure \ref{fig:randErr1pcdQvsdNdz} in particular, we see instability for three charge densities, which are not yet locked onto the stop-band. Whether a stop-band extends above or below a half-integer tune depends on it being a result of an effective focusing or defocusing error. Additionally, the stop-band itself is displaced with respect to the lattice tune-shift, caused by the quadrupole error.

We increase the magnitude of the random errors in order to find the upper limit of the lattice stability. The lattice becomes unstable with random error variations higher than $\pm$\SI{3}{\percent} for all tested seeds. Using $\pm$\SI{3}{\percent} random errors with the same seed like in the previous simulation case, we obtain more charge densities shifting the envelope tune into the stop-band and their eigenvalue moduli slightly increase as well. This is also true for their corresponding coupling instabilities in the other transverse plane. Furthermore, the charge densities necessary to shift the envelope tune into the stop-band has increased, since the lattice tune has shifted more due to the stronger error.

In order to understand how the beam size increased when the beam crosses a stop-band, we analyze how stop-bands lead to instability and subsequent envelope blow-up. We first find the fixed-point of a beam in an undisturbed lattice. The periodic beam envelope is parametrized as

\begin{equation}
\sigma_{0} = \mathcal{A}_{0}^{-1}\begin{bmatrix}\epsilon_{0} & 0\\ 0& \epsilon_{0}\end{bmatrix}\left(\mathcal{A}_{0}^{-1}\right)^{T} \quad \mathrm{with} \quad \mathcal{A}_{0} = \begin{bmatrix}\frac{1}{\sqrt{\beta}_{0}} &0\\ \frac{\alpha_{0}}{\sqrt{\beta}_{0}}& \sqrt{\beta}_{0}\end{bmatrix},
\label{eq:sig}
\end{equation}
where $\epsilon_{0}$ is the beam emittance and $\mathcal{A}_{0}$ contains the respective Twiss-parameters. The stop-band is created by adding a quadrupole $Q$ to a rotation matrix $\mathcal{O}$

\begin{equation}
Q = \begin{bmatrix}1 & 0\\ \frac{1}{f} & 1\end{bmatrix}; \quad \mathcal{O} = \begin{bmatrix}\cos{\mu} &\sin{\mu}\\ -\sin{\mu} & \cos{\mu} \end{bmatrix},
\label{eq:QO}
\end{equation}
with focal length $f$ and phase-advance $\mu$. The full-turn transfer matrix is given by $R=\:Q\mathcal{O}\left(\mu\right)$. Matrix $R$ can be parametrized by $R=\: \mathcal{A}^{-1}\mathcal{O}\left(\hat{\mu}\right)\mathcal{A}$, where $\mathcal{O}\left(\hat{\mu}\right)$ is a pure rotation and matrix $\mathcal{A}$ contains the Twiss-parameters. $\hat{\mu}$ is the phase-advance, shifted by the quadrupole error and is approximately $\hat{\mu}\approx \mu-\beta/4\pi f$.

\begin{figure}[bt]
        \centering
        \includegraphics[width=0.6\textwidth]{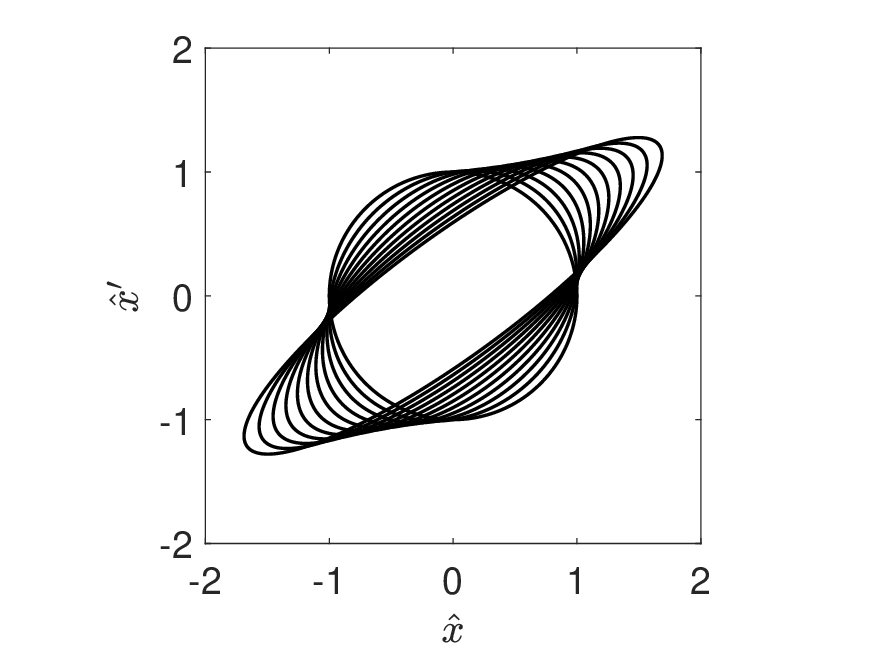}
 \caption{Stretched phase-ellipse due to imaginary phase-advances and Twiss-parameters during stop-band crossing.}
		\label{fig:deformEllipse}
\end{figure}

If the undisturbed phase-advance $\mu$ is close to zero or \num{2}$\pi$, the tune-shift of the quadrupole errors moves the beam onto the integer stop-band and the decomposition of the transfer matrix $R$ causes the matrix $\mathcal{A}$ to become complex, which we denote by $\tilde{\mathcal{A}}$. Furthermore, the decomposition yields a rotation matrix $\mathcal{O}\left(\tilde{\mu}\right)$ with a now complex phase-advance $\tilde{\mu}=\: 0 + \mathrm{i}m$, where the real part is locked on zero. The zero real part of the rotation yields the unit matrix and can thus be omitted. We perform the beam propagation $\sigma= R\:\sigma_{0}\:R^{T}$ in the stop-band, which in re-composed parametrization reads

\begin{equation}
\sigma=\epsilon_{0} \mathcal{\tilde{A}}^{-1}\mathcal{O}\left(\mathrm{i}m\right)\mathcal{\tilde{A}} \mathcal{A}_{0}^{-1} \left(\mathcal{A}_{0}^{-1}\right)^{T} \mathcal{\tilde{A}}^{T} \mathcal{O}\left(\mathrm{i}m\right)^{T} \left(\mathcal{\tilde{A}}^{-1}\right)^{T}.
\label{eq:sigma1full}
\end{equation}
During the matrix multiplication, all involved imaginary parts create real contributions, which distorts the circle in phase-space into an ellipse, thus inducing a mismatch to the beam. Figure \ref{fig:deformEllipse} shows the deformation of a normalized phase-ellipse for ten subsequent turns through a stop-band. This deformation is the envelope blow-up which is observed during stop-band crossing.

\FloatBarrier
\subsection{Envelope Studies with single and random skew Quadrupoles}

\begin{figure}[bt]
    \centering
    \begin{subfigure}{0.49\textwidth}
        \centering
        \includegraphics[width=\textwidth]{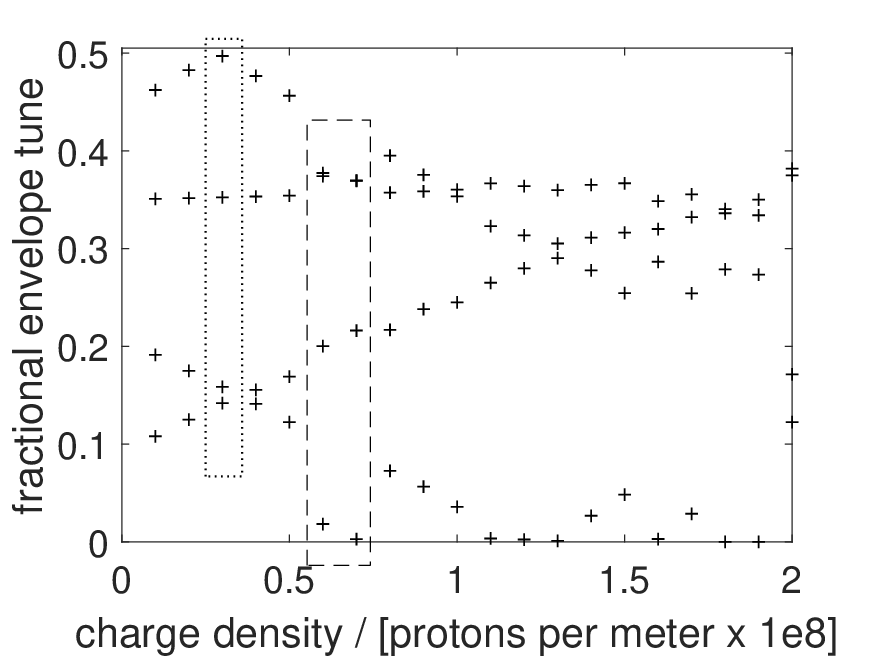}
        \caption{}
				\label{fig:oneSkew1degSet2Eig}
    \end{subfigure}%
 ~
		    \begin{subfigure}{0.49\textwidth}
        \centering
        \includegraphics[width=\textwidth]{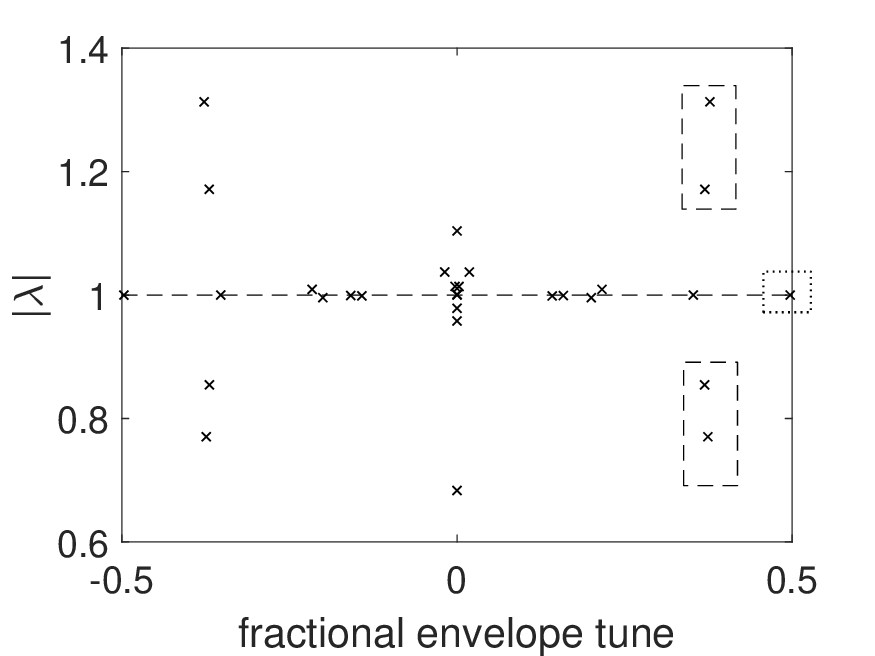}
        \caption{}
				\label{fig:oneSkew1degSet1Eig}
    \end{subfigure}%
    \caption{a) Coherent tune-shift for all four eigenmodes, where a single quadrupole in the lattice received a \SI{1}{\degree} roll angle. Coupling resonance, where a single mode exhibits a flip-flop behavior (dotted box), and coupling resonances due to at least two modes obtaining nearly identical tunes (dashed box). b) Moduli of the eigenvalues corresponding to the discussed coupling resonances.}
		\label{fig:oneSkew1degEig}
\end{figure}

We now turn to the effect of transverse coupling in the magnet lattice and model this by rolling one focusing quadrupole by \SI{1}{\degree}. This causes the periodic beam matrix---the fixed point---to acquire non-zero values for the components describing transverse coupling. Moreover, the stability analysis is now based on finding the eigenvalues of a $10\times 10$-matrix, which causes two additional pairs of complex-conjugate eigenvalues to appear. We show the simulation results in Fig. \ref{fig:oneSkew1degEig} and focus on distinct cases of coupling instabilities, which are introduced by the rolled quadrupole.

\begin{figure}[bt]
    \centering
    \begin{subfigure}{0.49\textwidth}
        \centering
        \includegraphics[width=\textwidth]{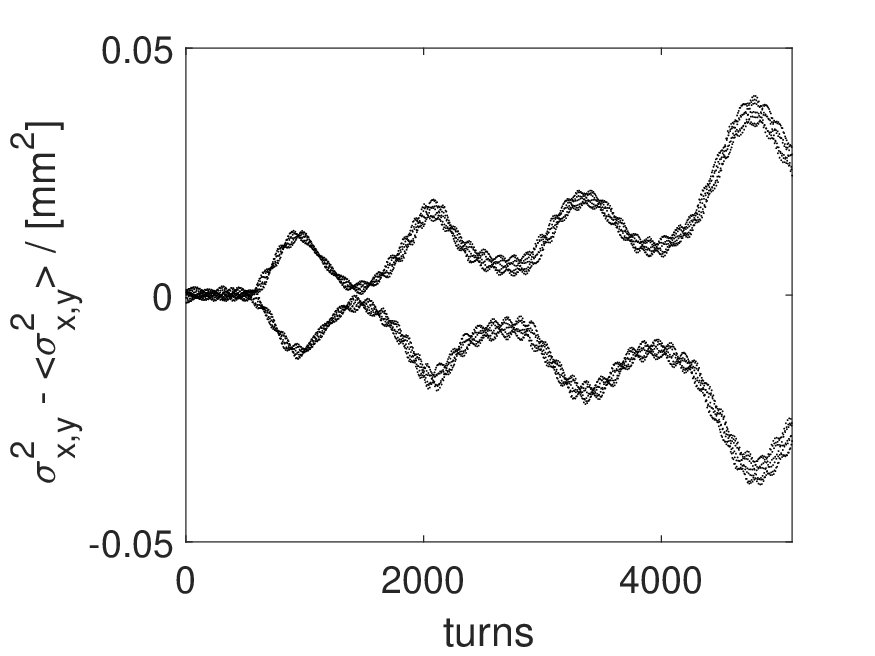}
        \caption{}
				\label{fig:3e7Resonset}
    \end{subfigure}%
 ~
		    \begin{subfigure}{0.49\textwidth}
        \centering
        \includegraphics[width=\textwidth]{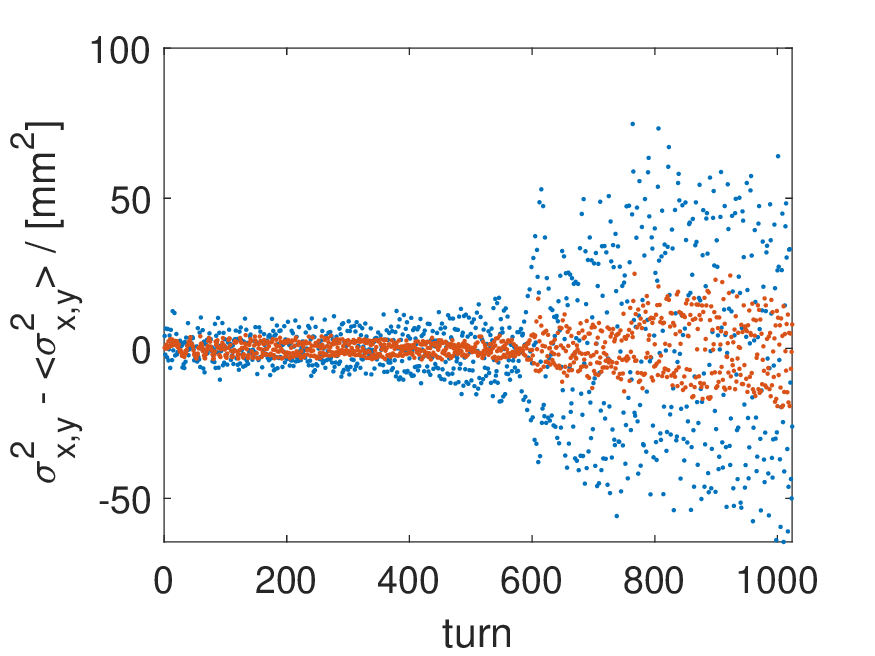}
        \caption{}
				\label{fig:res6}
    \end{subfigure}%
    \caption{a) Horizontal beam envelope in a flip-flop mode for a charge density of \num{0.3e8} protons per meter.  b) Horizontal (blue) and vertical (red) beam envelopes in strong coupling resonance for charge densities of \numlist{0.6e8;0.7e8} protons per meter, where at least two eigenmodes exhibit nearly identical tunes.}
		\label{fig:3e7Res}
\end{figure}

The first example of a coupling instability occurs at a charge density of \num{0.3e8} protons per meter (dotted rectangle). One of the four eigenvalues lies on a half-integer tune, its complex-conjugated partner on the zero fractional tune. This indicates that the mode exhibits a flip-flop behavior. The corresponding envelope evolution is shown in Fig. \ref{fig:3e7Resonset}. The beam size jumps between the upper and lower path with each turn while having a small, but steady growth and confirms the mode as a resonant flip-flop mode.

\noindent

We observe another coupling instability for charge densities of \numlist{0.6e8;0.7e8} protons per meter (dashed box). In this example, the fractional tunes of two modes are very close to each other, resulting in a difference resonance. All four eigenvalues show instability, but the two resonant modes have significantly larger moduli. The corresponding envelope evolution is shown in Fig. \ref{fig:res6} and shows a large envelope growth over only \num{1024} turns. Further resonances of this type occur at charge densities of \numlist{1.3e8;1.8e8} protons per meter, where at least two of the four modes obtain a nearly identical tune. The presence of four eigenmodes --- together with space-charge tune-shift and changes in lattice tune --- opens for more possibilities to fulfill a resonance condition. Since the strength of a rolled quadrupole deviates from its nominal value in the projected transverse planes, it produces small integer stop-band contributions. This can be observed for charge densities between \numrange{1.1e8}{1.3e8} protons per meter, where the tunes are locking onto the integer.

Introducing random roll angles to {\em all} quadrupoles along the lattice changes the locations at which coupling resonances appear. Like with random quadrupole errors, the lattice coupling may compensate, depending on the seed. We find that the lattice becomes unstable, if the variation on the random roll angles exceeds $\pm$\SI{1}{\degree}.

\FloatBarrier
\subsection{Tune-shift of mismatched Beams}
\label{subsec:mismatch}

So far, we examined envelope stability by slightly perturbing the periodic beam solution and found that crossing resonances causes the mismatch to grow. We therefore investigate the effect of a large mismatch and examine how the envelope tune-shift changes with the mismatch. We quantify mismatch by the factor $B_{mag}$ \cite{Raubenheimer:1995vh}, which describes the amount of non-overlapping between two ellipses in phase space and which is given by

\begin{equation}
B_{mag} = \frac{1}{2}\left[ \frac{\beta^{*}}{\beta} + \frac{\beta}{\beta^{*}} + \left( \alpha\sqrt{\frac{\beta^{*}}{\beta}}-\alpha^{*}\sqrt{\frac{\beta}{\beta^{*}}} \right)^{2} \right],
\end{equation}
where $\alpha$ and $\beta$ are the Twiss-parameters of a periodic solution and the starred Twiss-parameters those of the mismatched beam . We apply the mismatch by manipulating the $\sigma_{11}$ and $\sigma_{22}$ elements of the periodic beam matrix such that the emittance is preserved.

\begin{figure}[hbt]
    \centering
\begin{subfigure}{0.49\textwidth}
        \centering
        \includegraphics[width=\textwidth]{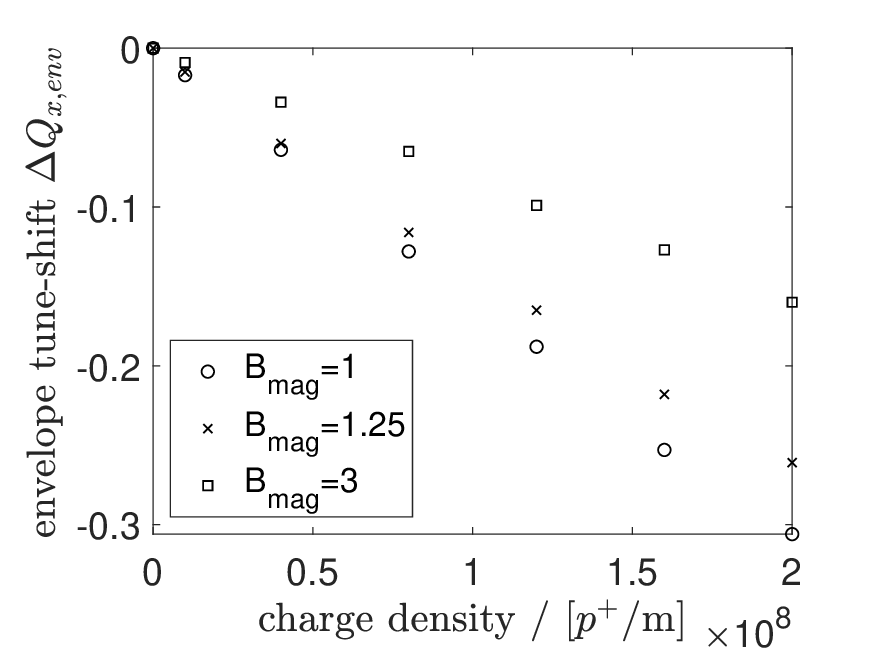}
				\caption{}
		\label{fig:tuneshifts}
    \end{subfigure}%
		~
		    \centering
    \begin{subfigure}{0.49\textwidth}
        \centering
        \includegraphics[width=\textwidth]{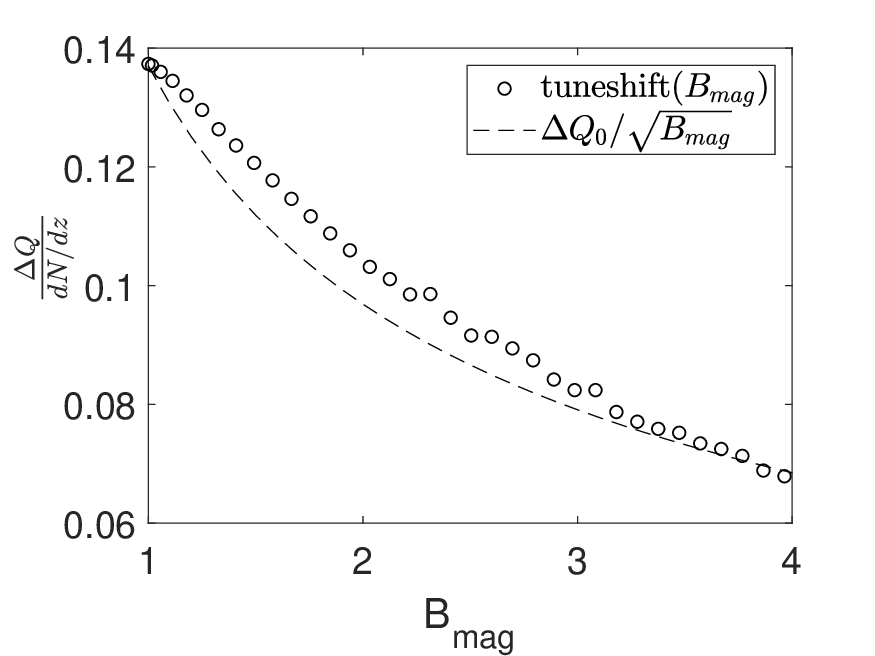}
        \caption{}
				\label{fig:TSvsBmag}
    \end{subfigure}%
    \caption{a) Coherent envelope tune-shift as a function of charge density for different mismatch factors $B_{mag}$. Envelope tune-shift decreases with increasing beam mismatch. b) Tune-shift rate as a function of $B_{mag}$.}
		\label{fig:TSvsCurr}
\end{figure}

Figure \ref{fig:tuneshifts} shows the envelope tune-shifts for a matched beam and two mismatched beams, obtained from Fourier analysis of tracked envelopes, as function of charge density. The tune-shift increases linearly with the current for $B_{mag}=\:$1, whereas the slope becomes flatter for increasing beam mismatch. The absolute slope coefficients of the linear tune-shift as a function of beam mismatch is shown in figure \ref{fig:TSvsBmag}. The linear tune-shift is proportional to

\begin{equation}
\Delta Q\propto \frac{1}{\sigma_{x}\sigma_{y}} \propto  \frac{1}{\sqrt{B_{mag}\beta_{x,0}\beta_{y}}}.
\end{equation}
Since all mismatched beams stem from the same matched beam sizes, we omit the dependency on the beta-functions such that the dependency on $B_{mag}$ is well-approximated by $\Delta Q\approx\:\frac{\Delta Q_{0}}{\sqrt{B_{mag}}}$. The approximation agrees well to the simulation results as seen in Figure \ref{fig:TSvsBmag}. The exact dependency of the tune-shift on beam mismatch varies since mismatch changes the beta-function of the envelope and thus, the integrated space-charge force which itself couples back into a change in beta-function and tune.

\section{Conclusions}
\label{sec:conc}

We presented a non-linear space-charge model and examined the beam envelope dynamics and stability for different lattice configurations including quadrupole errors and skew quadrupoles by means of eigenvalue analysis and as a function of charge density.

First, we tested the principle of eigenvalue analysis on a single cell where we compared its envelope tunes to those obtained by Fourier analysis from tracked envelopes. Both methods agree well, showing that the eigenvalue analysis yields correct envelope tunes. In simulations with the complete ring, we found that a  mode, crossing a zero fractional tune, exhibits instabilities, despite the absence of lattice errors. We explain this by interpreting the space-charge force as defocusing quadrupole error, which creates a stop band at zero fractional tunes. The main qualitative difference between our model and approaches with single particle tracking is that any form of instability leads to a growing envelope whereas single particles would either be lost during instability or rearrange themselves, leading to different rms quantities. Furthermore, we found higher order, space-charge driven resonances.

Applying a single quadrupole error creates stop-bands. In this case, more eigenvalues lie in the stop-band because its width accepts a broader range of shifted envelope tunes. We found, that whenever an envelope is in the stop-band and becomes strongly unstable, the other transverse envelope is affected as well, albeit with weaker instability. This is attributed to the intrinsic coupling between the transverse planes via the space-charge force.

Introducing random quadrupole errors did not lead to broader stop-bands. However, the exact width of the stop-band depends on the error seed. For the same reason, triggered stop-band instabilities and coupling resonances change because they are the result of the combination between coherent envelope tune-shift and error-induced shift in the lattice tune. Increasing the strength of the error within the same seed led to more unstable eigenvalues with larger moduli. We found the stability limit of the lattice with random error variation between $\pm$\SI{3}{\percent} for many tested seeds. Beyond that, the tune of the lattice becomes imaginary and thus, inherently unstable.

We presented an analysis of stop-band-induced instabilities and showed that the initially circular phase-ellipse is stretched during stop-band crossing. This deformation is the result of an imaginary phase-advance in the stop-band, which causes a mismatch that leads to an increased beam size.

Our model enables us to treat coupled lattices, where \num{4} eigenmodes needed to be observed. We have presented two examples of coupling resonances, which appeared after giving a roll angle of \SI{1}{\degree} to a single focusing quadrupole. In the first example, we found an unstable envelope mode with a flip-flop behavior. This behavior is indicated by an eigenvalue pair lying on the tunes \num{0} and \num{0.5}, respectively. The unstable flip-flop behavior was confirmed by envelope tracking. Secondly, we found strong coupling resonances whenever two modes exhibit nearly identical tunes. The respective eigenvalue moduli clearly increased and the envelope tracking confirms simultaneously growing envelopes with strong growth already in a short time-frame. Introducing random roll angles to all quadrupoles changes occurrences of coupling resonances and depend on the individual seed. We found that the lattice becomes unstable if the variation on the random roll angles exceeds $\pm$\SI{1}{\degree}.

Additionally, we examined the effect of beam mismatch on the linear tunes-shift by manipulating the $\sigma_{11}$ and $\sigma_{22}$ elements of the periodic beam matrix. We found that the linear tune-shift becomes weaker with larger mismatch and we showed with a simple proportionality, that the ratio between the mismatched and the matched tune-shift is well-approximated by $\frac{\Delta Q_{mismatch}}{\Delta Q_{0}}\approx\:\frac{1}{\sqrt{B_{mag}}}$.

\section{Acknowledgements}
This research did not receive any specific grant from funding agencies in the public, commercial, or not-for-profit sectors.

\bibliography{ms}

\end{document}